\documentclass[aps,prl,superscriptaddress,floatfix,onecolumn, notitlepage, preprint]{revtex4-1}

\usepackage{amsmath,amssymb,amsfonts}
\usepackage{hyperref}
\usepackage{graphicx}
\usepackage{color}
\usepackage{gensymb}
\usepackage{LatexCommands}

\definecolor{myblue}{rgb}{0, 0.33, 0.83}

\begin{document}

\title{Visualizing band selective enhancement of quasiparticle lifetime in a metallic ferromagnet}

\author{Na Hyun Jo}
\thanks{These two authors contributed equally}
\author{Yun Wu}
\thanks{These two authors contributed equally}
\author{Tha\'{i}s V. Trevisan}
\affiliation{Division of Materials Science and Engineering, Ames Laboratory, Ames, Iowa 50011, USA}
\affiliation{Department of Physics and Astronomy, Iowa State University, Ames, Iowa 50011, USA}

\author{Lin-Lin Wang}
\affiliation{Division of Materials Science and Engineering, Ames Laboratory, Ames, Iowa 50011, USA}

\author{Kyungchan Lee}

\author{Brinda Kuthanazhi}

\author{Benjamin Schrunk}

\author{S.~L.~Bud'ko}

\author{P. C. Canfield}

\author{P. P. Orth}

\author{Adam Kaminski}
\email[]{kaminski@ameslab.gov}
\affiliation{Division of Materials Science and Engineering, Ames Laboratory, Ames, Iowa 50011, USA}
\affiliation{Department of Physics and Astronomy, Iowa State University, Ames, Iowa 50011, USA}

\date{\today}

\maketitle

\textbf{
Electrons navigate more easily in a background of ordered magnetic moments than around randomly oriented ones. This fundamental quantum mechanical principle is due to their Bloch wave nature and also underlies ballistic electronic motion in a perfect crystal. As a result, a paramagnetic metal that develops ferromagnetic order often experiences a sharp drop in the resistivity. Despite the universality of this phenomenon, a direct observation of the impact of ferromagnetic order on the electronic quasiparticles in a magnetic metal is still lacking. Here we demonstrate that quasiparticles experience a significant enhancement of their lifetime in the ferromagnetic state of the low-density magnetic semimetal EuCd$_2$As$_2$, but this occurs only in selected bands and specific energy ranges. This is a direct consequence of the magnetically induced band splitting and the multi-orbital nature of the material. Our detailed study allows to disentangle different electronic scattering mechanisms due to non-magnetic disorder and magnon exchange.
Such high momentum and energy dependence quasiparticle lifetime enhancement can lead to spin selective transport and potential spintronic applications.
}
The development of long-range electronic order has a strong impact on electronic structure and transport properties of quantum materials. For charge density wave (CDW) and spin density wave (SDW) orders that are characterized by a finite-$q$ wave vector, this is a result of band folding and Fermi surface reconstruction due to new (larger) periodicity in the crystal. This has been observed in number of systems including NbSe$_{2}$\,\cite{Borisenko2009}, Fe pnictide\,\cite{liu2010,Kondo2010}, CrAuTe$_{4}$\,\cite{Jo2016} and CeSb\, \cite{kuroda2020}. In contrast, for ferromagnetic (FM) order the periodicity remains unchanged, but instead the bands experience a FM exchange splitting, leading to different Fermi surfaces for majority and minority carriers. The impact of such band splitting and coupling to ordered moments on itinerant carrier properties in complex materials is an important open question that we address here. Previous studies have focused on itinerant magnets with 3$d$ transition metals, where electronic bands are broad due to the strongly correlated nature and the observed effects on quasiparticle properties are small. The phenomenon of FM band splitting was demonstrated, for example, in elemental Ni\,\cite{Eastman1978}, and a more recent ARPES study on Fe$_{3}$GeTe$_{2}$ also revealed FM exchange splitting with reduced quasiparticle coherence\,\cite{XuLi2020}. Rare-earth-based FM materials with local magnetic moments and weakly correlated conduction electron bands are a much better platform for studying the impact of ferromagnetic order on electronic quasiparticle properties.
Here, we investigate a ferromagnetic variant of EuCd$_{2}$As$_{2}$, which has received much attention recently as a candidate magnetic Weyl semimetal~\cite{Hua2018Dirac,Wang2019Single, Soh2019Ideal,ma2019} displaying ferromagnetic~\cite{Jo2019,taddei2020spin} or antiferromagnetic order~\cite{Artmann1996AM2X2, Schellenberg2011Mossbauer,Rahn2018Coupling}, which can be controlled by doping.
These unique properties make FM-EuCd$_{2}$As$_{2}$ an ideal candidate for such studies. We show that its electronic quasiparticles experience a significant lifetime enhancement in certain bands and energy ranges. We associate this effect with the emergence of different impurity and magnetic scattering rates for majority and minority carriers due to distinct phase spaces. This demonstrates the complexity and importance of magnetic coupling to itinerant carriers and establishes a direct connection between quasiparticle properties and transport behavior in metallic FMs.
\begin{figure*}[tb]
	\includegraphics[width=6 in]{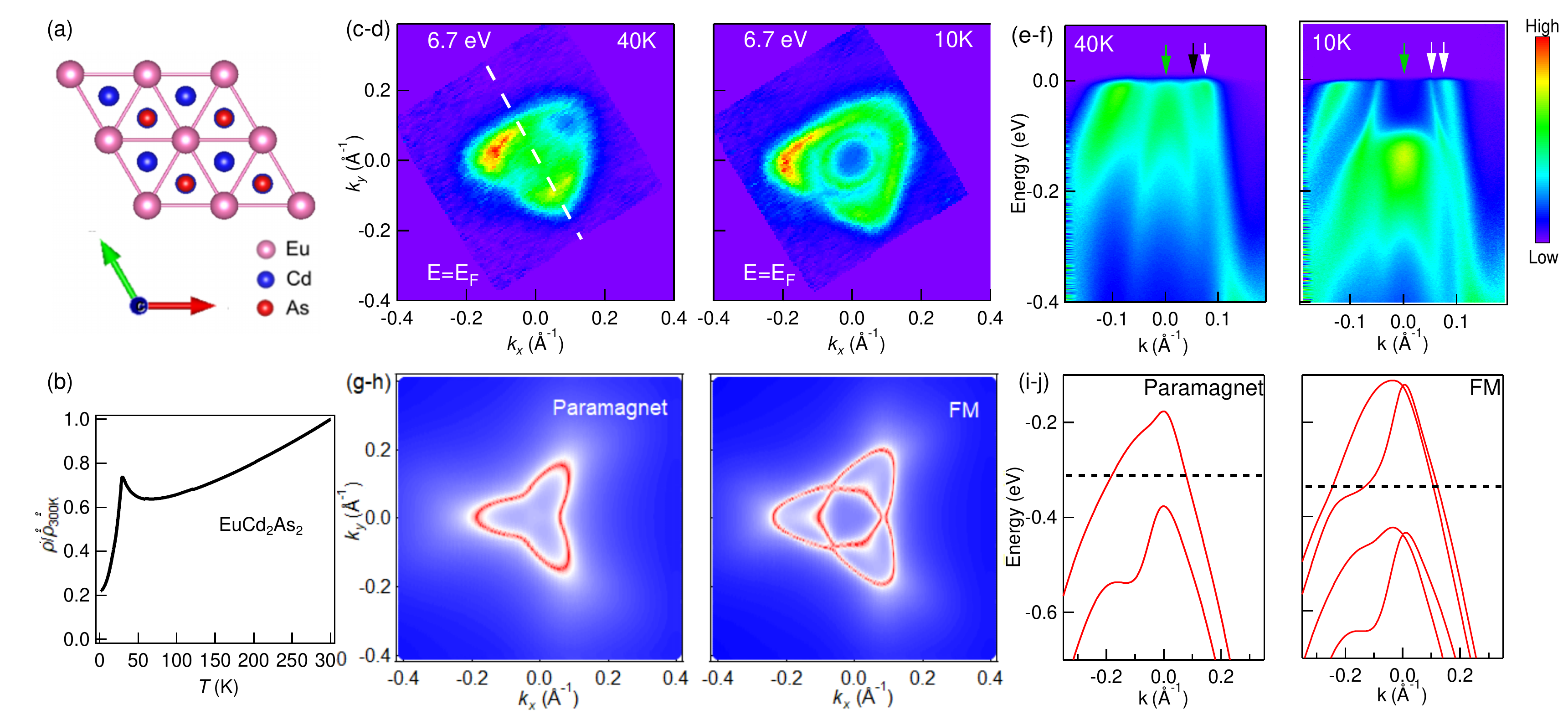}%
	\caption{\textbf{Electronic structure and Fermi surface cut from ARPES and ab initio theory in para- and ferromagnetic phase, crystal structure and resistivity}
	(a) Crystal structure of EuCd$_2$As$_2$.
	(b) Temperature dependent normalized resistivity, $\rho(T)/\rho(300~\text{K})$, where $\rho(300~\text{K})\,\sim\,4\,\times\,10^{-4}\,\ohm$\,cm.
	(c-d) Fermi surface results from ARPES measurements at 40~K and 10~K, respectively, where $T_{\text{C}} = 26$~K is ferromagnetic transition temperature.
	(e-f) ARPES intensity along the white dashed lines in (a-b) measured at 40~K and 10~K, respectively. The white arrows point to the bands crossing the Fermi level; the green arrow points to the fully occupied hole band at the center; the black arrow points to a possible splitting of the bands at 40~K.
	(g-h) Fermi surface calculated using DFT in para- and ferromagnetic states, respectively.
	(i-j) Band dispersion from DFT calculations along the white dashed line in (c).
	\label{fig:Fig1}}
\end{figure*}

Figure~\ref{fig:Fig1} presents the crystal structure, temperature dependent resistance, Fermi surface cut and band dispersion of EuCd$_{2}$As$_{2}$ obtained using ARPES and DFT calculations. Temperature dependent normalized resistivity data (Fig.\,\ref{fig:Fig1} (b)) clearly show a rapid drop of resistance below $T_{\textrm{C}}$ as a result of the loss-of-spin-disorder scattering. An upturn above the transition temperature is likely associated with magnetic fluctuations. Fig.~\ref{fig:Fig1} (c) displays the Fermi surface cut of EuCd$_{2}$As$_{2}$ measured at 40~K (i.e., above $T_{\textrm{C}}$\,$\sim$\,26 K), where an outer Fermi surface sheet with three-fold rotation symmetry and some intensity at the center of the Brillouin zone can be clearly seen. The symmetry of the outer Fermi surface sheet above $T_{\textrm{C}}$ is consistent with the three-fold symmetry of the crystal structure in $ab$ plane as shown in Fig.\,\ref{fig:Fig1} (a). As the sample is cooled down to 10~K, i.e., below the ferromagnetic transition temperature\,\cite{Jo2019}, the outer trigonal Fermi surface sheet expands a little bit. What is more astonishing is that a sharp circular hole pocket emerges at the center of the Brillouin zone as shown in Fig.~\ref{fig:Fig1} (d). This is clearly the result of the ferromagnetic transition\,\cite{Jo2019}, which leads to the majority and minority spin splitting in this unique ferromagnetic, semimetallic system. This phenomena is better visualized in Figs.~\ref{fig:Fig1} (e-f), that show the band dispersion along the white dashed lines in (c). The white arrows point to the hole bands crossing the Fermi level, forming the circular (inner) and trigonal (outer) Fermi surface sheets seen in panels (c-d). The green arrow points to the center hole band that sinks down dramatically to roughly 100~meV below Fermi level as the EuCd$_{2}$As$_{2}$ sample undergoes ferromagnetic transition. The black arrow in Fig.~\ref{fig:Fig1}(e) points to the possible band splitting of EuCd$_{2}$As$_{2}$ in paramagnetic state. (Details will be discussed later.) Figs.~\ref{fig:Fig1}(g-h) show the DFT calculated Fermi surfaces of EuCd$_{2}$As$_{2}$ in the paramagnetic and ferromagnetic states ($k_{z}$ = 0.17 ($\pi/c$)), which match relatively well with the ARPES results shown above. Note that, in order to achieve a better agreement with the ARPES results presented in Figs.~\ref{fig:Fig1}(c-d), we have to shift the chemical potential of the calculation results downward by roughly 450~meV for both the paramagnetic and ordered state. This is also consistent with presence of Eu vacancies suggested by the powder X-ray data~\cite{Jo2019}. Figs.~\ref{fig:Fig1}(i-j) show the band dispersion of EuCd$_{2}$As$_{2}$ calculated from DFT, showing reasonable agreement with the ARPES measurements.

\begin{figure*}[tb]
	\includegraphics[width=6 in]{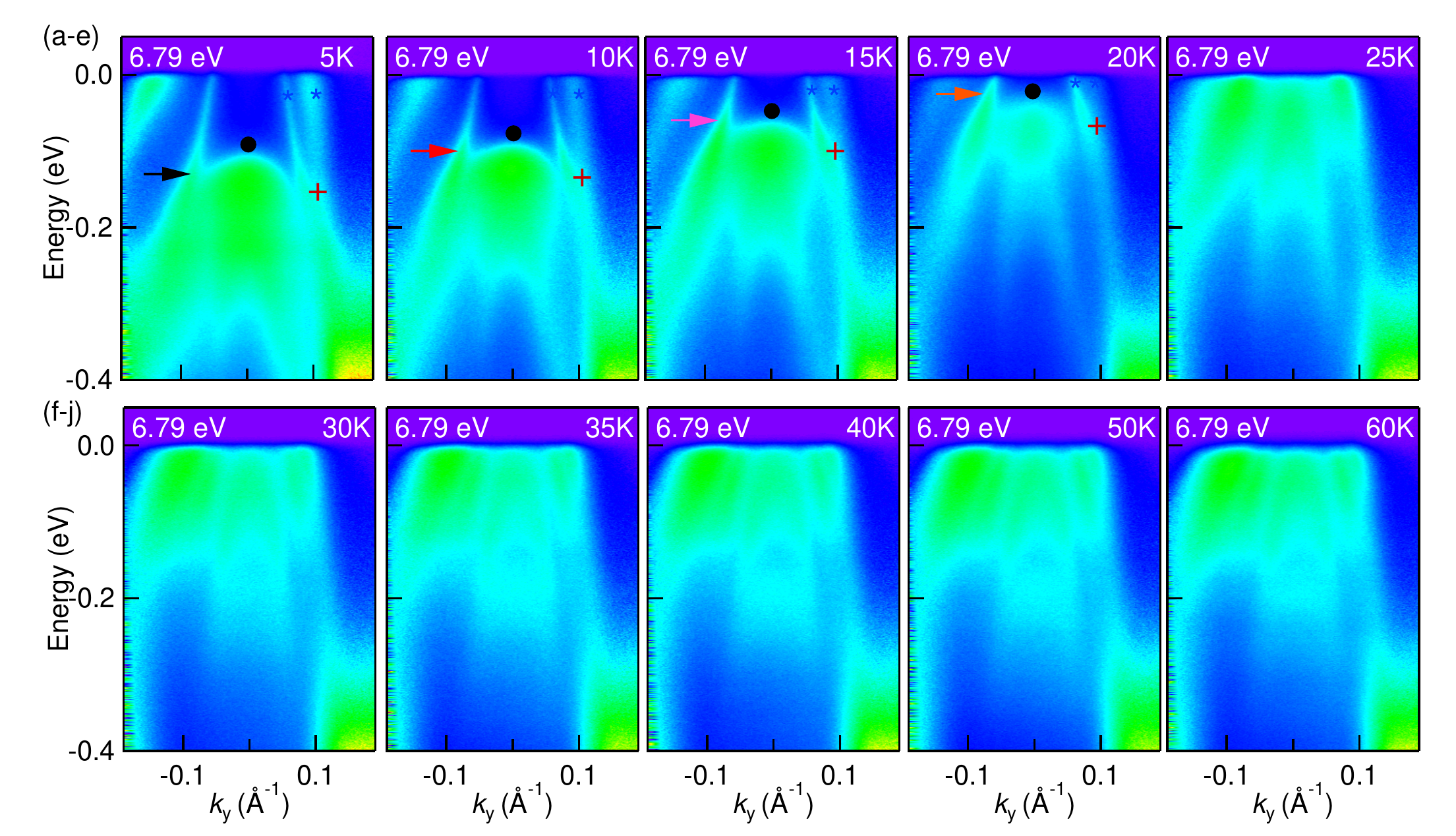}%
	\caption{\textbf{Temperature evolution of electronic band dispersion in EuCd$_{2}$As$_{2}$.}
	(a-j) Band dispersion of measured using 6.79~eV photons at indicated temperatures between $5$~K and $60$~K. The arrows mark the energy above which enhancement of quasiparticle lifetime occurs. Black filled circles denote the top of an inner hole band, which continuously drops in energy below $T_{\text{C}}$. Blue stars point out the splitting of hole bands occurs on the right side. Red crosses mark the energy at which the splitting of hole bands occurs on the right side of $\Gamma$ point.
	\label{fig:Fig2}}
\end{figure*}

In addition to band splitting we also observed a large enhancement of quasiparticle lifetime that accompany the magnetic transition. This can be quite clearly seen in raw data in Figs.~\ref{fig:Fig1}(e-f), where very sharp band is present only below $T_{\textrm{C}}$. In Figs.~\ref{fig:Fig2}(a-j) we show the detailed temperature evolution of the band dispersion measured using 6.79 eV photons at select temperatures between 5~K and 60~K. The results were reproduced using several samples and temperature cycling. With decreasing temperature, the sample undergoes a ferromagnetic transition around 26K. Two significant features can be observed. One is the enhancement of the quasparticle lifetime at the inner hole pocket, and the other is the downward shift of the fully occupied band centered at $\Gamma$ with the decreasing  temperature below $T_{\textrm{C}}$. It is very interesting that the enhancement of quasiparticle lifetime occurs predominantly for the the inner hole band and only over limited energy range that changes with temperature. This can be seen in Fig.\,\ref{fig:Fig2}~(a-d), where the  hole band forming the inner circular Fermi pocket is sharp until it reaches the binding energy that roughly corresponds to the top of the fully occupied center hole band (indicated by the red/black arrows). This effect suggests a significant interband electron-electron scattering between the inner hole band and the center hole band, while intraband electron-electron scattering is strongly suppressed below $T_{\textrm{C}}$.

\begin{figure*}[tb]
	\includegraphics[width=6.5 in]{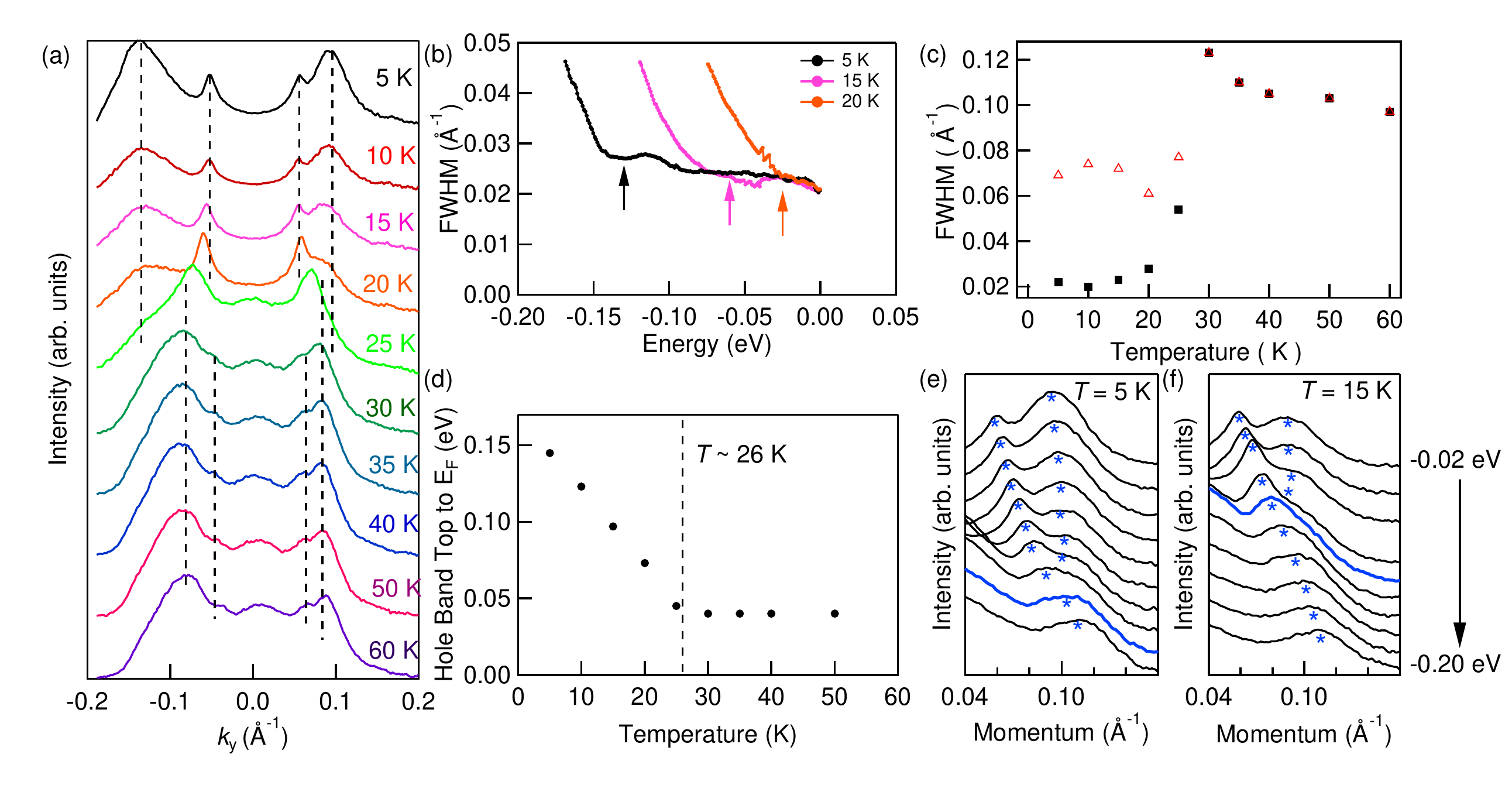}%
	\caption{\textbf{Detailed temperature evolution of the electronic structure of EuCd$_{2}$As$_{2}$ obtained using ARPES.}
	(a) MDCs at the Fermi energy measured at temperatures between 5~K and 60~K. The dashed lines are guide to an eye marking locations MDC peaks.
	(b) Full Width Half Maximum (FWHM) data obtained from Lorentzian fits to the MDCs for the inner hole band showing significant reduction of scattering (enhancement of lifetime) for energies slightly above the top of the fully occupied band, which are marked by arrows.
	(c) FWHM of the MDC peaks on the left side in Fig.\,\ref{fig:Fig3} (a) as a function of temperature. Red open triangles represent the very left peak (broad outer band), and black filled square represent the second left peak (sharp inner band).
	(d) The binding energy of the fully occupied hole band top as a function of temperature.
	(e-f) MDCs at 5\,K and 15\,K, respectively, for the right side of hole bands close to energies marked as blue stars in Fig.\,\ref{fig:Fig2}; -0.02~eV to -0.20~eV with 0.02~eV steps. Blue stars indicate the peak position. MDCs for which the two peaks merge are marked in blue.
	\label{fig:Fig3}}
\end{figure*}

To better visualize the band splitting and suppression of the scattering effect at the FM transition, we performed a detailed analysis summarized in Figure\,\ref{fig:Fig3}. Panel (a) shows Momentum Distribution Curves (MDCs) at the Fermi energy measured at various temperatures. The changes in the electronic structure at the magnetic transition temperature of $\sim$\,26\,K are quite prominent. Two well separated bands at low temperatures merge into one broad band above the transition temperature. In addition, broad peak appears at k$_y$=0 due to upward shift of the fully occupied center band. To demonstrated quantitatively the enhancement of quasiparticle lifetime, we plotted the Full Width Half Maximum (FWHM) of the inner hole band extracted by using Lorentzian fitting to MDC in Fig.~\ref{fig:Fig3}~(b). We can clearly see that the FWHM is small and stays relatively constant between E$_F$ and binding energy indicated by arrows for each temperature. It then increases dramatically for higher binding energies, demonstrating the unusual enhancement of the quasiparticle lifetime. We also plotted the FWHM of the left side of bands, extracted from multi-peak Lorentzian fits to MDCs (shown in the Fig.\,\ref{fig:Fig3} (a)), as a function of temperature in panel c. The red open triangles represent the broad outer band, and the black filled squares represent the inner sharp band. Interestingly, the graph somewhat resembles the temperature dependent resistance graph, shown in Fig.\,\ref{fig:Fig1} (b), especially for the sharp band: slight upturn above the transition temperature followed by rapid decrease of FWHM below the $T_{\textrm{C}}$. Although both outer and inner bands experience sharpening of the band, the effect is about three times larger for the inner band. If we compare the data from inner band at 5\,K to 30\,K data, the FWHM is about six times smaller. The other interesting feature is the continuous downward shift of the center hole band with the decreasing temperature below the magnetic transition. As we can see from Fig.\,\ref{fig:Fig2}, the top of the center hole band almost touches the Fermi level and stays at the same energy above 26~K. As the sample temperature decreases, the center hole band continuously moves downward and reaches around 150~meV below Fermi level at 5~K. The energy location of the top of the center hole band seems to follow an order parameter like behavior of the magnetic ordering in EuCd$_{2}$As$_{2}$. By extracting the peak positions of the Energy distribution curves, i.e., the top of the center hole band, we plotted the result in Fig.\,\ref{fig:Fig3} (d) showing a clear transition and enhancement of the Zeeman splitting that is a response to increasing internal field (i.e., FM order parameter) as the temperature is decreased. Figure\,\ref{fig:Fig3} (d-e) show MDCs for the right side of the hole bands marked with blue stars in Fig.\,\ref{fig:Fig2}. A single band splits into two components below the transition temperature. Furthermore, the splitting point of two bands gets close to $E_{\textrm{F}}$ as temperature approaches to the transition temperature.

\begin{figure}[tb]
	\centering
	\includegraphics[width=\linewidth]{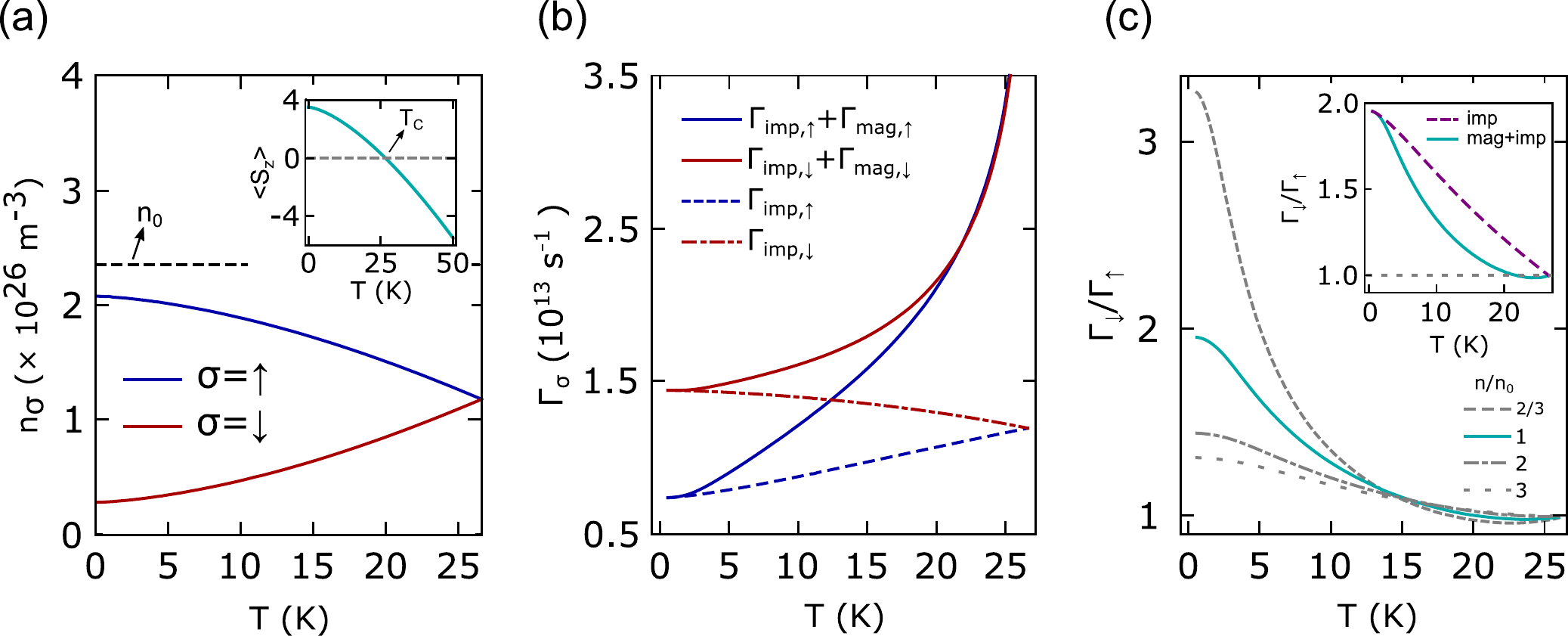}
	\caption{\textbf{Theoretical model results.} (a) Spin resolved carrier density $n_\sigma$ ($\sigma = \uparrow, \downarrow$) as a function of temperature $T$ below the magnetic transition $T_{\text{C}}  \approx 26$~K in the theoretical model. The total carrier density $n_0 = n_\uparrow + n_\downarrow = 2.35 \times 10^{26}$~m$^{-3}$ is taken from Hall measurements. The other parameters are set to $J S \hbar^2= 150$~meV, $J_0 S \hbar^2= 0.15$~meV, and $m^* = -m_e$ with bare electron mass $m_e$.
	Inset shows average $z$-component $\langle S_z \rangle(T)$ of Eu spins versus $T$, used to estimate FM coupling constant $J_0 S$ from $\langle S_z \rangle(T_{\text{C}}) = 0$.
	(b) Total minority and majority scattering rates $\Gamma_{\downarrow}$ and $\Gamma_{\uparrow}$ (solid) as a function of $T$, where $\Gamma_\sigma = \Gamma_{\text{imp},\sigma} + \Gamma_{\text{mag}, \sigma}$. Dashed lines show impurity scattering contributions, which dominate at low $T$. Upturn close to $T_{\text{C}}$ is caused by scattering with magnons that proliferate at the phase transition.
	(c) Ratio of minority over majority quasiparticle scattering rate, $\Gamma_\downarrow/\Gamma_\uparrow$, as a function of $T$ for different carrier densities $n/n_0$. The ratio is larger than unity except close to $T_{\text{C}}$, and increases for lower $T$ and $n$. Inset shows ratio obtained from total scattering rate (solid) and impurity contribution only (dashed) for $n = n_0$.
	}
	\label{fig:figure1_theory}
\end{figure}

\emph{Theoretical model.--}
To model the experimental observation of a band and energy selective enhancement of quasiparticle lifetimes in the ferromagnetically ordered state, we consider a system of itinerant electrons with hole-like dispersion
coupled to local Eu $S = 7/2$ magnetic moments (coupling constant $J$). The Eu moments interact ferromagnetically with each other with exchange constant $J_0$.
We also take potential disorder into account that arises, for example, from Eu vacancies~\cite{Jo2019} (see Methods section for details).
In the ordered state, the electronic bands Zeeman split into majority (minority) bands, where the electronic spin projection is mostly (anti-)parallel to the magnetization. This results in different densities $n_\sigma$ and Fermi surface volumes for majority ($\sigma = \uparrow$) and minority ($\sigma = \downarrow$) carriers with Fermi wave vector $k_{F,\uparrow} < k_{F, \downarrow}$ for hole pockets, as shown in Fig.~\ref{fig:figure1_theory}(a).
Here, we use $J S \hbar^2 \approx 150$~meV to recover the experimentally observed band splitting at $T = 5~\text{K}$ (see Fig.~\ref{fig:Fig1}). The splitting decreases at higher temperatures and vanishes at the FM transition temperature $T_\text{C} = 26$~K, which sets $J_0 S\hbar^2 = 0.15$~meV in our model calculation (see inset in Fig.~\ref{fig:figure1_theory}(a) and Methods section for details).

Band dependent quasiparticle lifetimes $\tau_\sigma$ then follow from the fact that the available phase space for scattering processes is different for majority and minority carriers. There are two important scattering channels for electrons in the FM phase, non-magnetic impurity scattering and magnetic scattering involving magnon exchange.
Interestingly, while majority carriers in a hole-like band experience a larger magnetic scattering rate, 
$\Gamma_{\text{mag}, \uparrow} > \Gamma_{\text{mag}, \downarrow}$, they experience less impurity scattering than minority carriers, $\Gamma_{\text{imp}, \uparrow} < \Gamma_{\text{imp}, \downarrow}$. Since impurity scattering dominates below the magnetic transition (except very close to $T_{\text{C}}$), the lifetime of the majority band is enhanced, $\tau_\uparrow > \tau_{\downarrow}$ at $T \ll T_{\text{C}}$, where $\tau_\sigma = 1/(2 \Gamma_\sigma)$ and $\Gamma_\sigma = \Gamma_{\text{imp}, \sigma} + \Gamma_{\text{mag}, \sigma}$.


Specifically, we calculate the electronic scattering rates $\Gamma_\sigma$ due to impurity and magnetic scattering within the second order Born approximation.
To lowest order in the electron-magnon coupling $J$, the electronic dispersion acquires a spin-dependent shift proportional to the $T$-dependent magnetization: $\xi_\bfk \rightarrow \xi_{\bfk,\sigma} - \sigma \gamma(T)$ with
$\gamma(T) = \frac{J\hbar^2}{2} \bigl[S-\frac{\zeta(\frac32) (k_B T)^{3/2}}{8 (J_0 S \hbar^2 \pi)^{3/2}} \bigr]$.
Here, we identify $\uparrow = +$, $\downarrow = -$, and $\zeta(x)$ is the Riemann zeta function. The second term accounts for the reduction of magnetization due to magnon excitations.
To second order in $J$, the magnetic scattering rate reads~\cite{woolseyElectronmagnonInteractionFerromagnetic1970}
\begin{align}
\Gamma_{\text{mag}, \mathbf{k}, \sigma} =\frac{\pi \mathcal{V} J^2\hbar^3 S }{2} \int_q \left[ \sigma\, n_F(\xi_{\mathbf{k}+ \sigma \mathbf{q},\bar{\sigma}})+n_{B}(\Omega_{\mathbf{q}}) + \delta_{\sigma, -1}\right]\delta\left(\xi_{\mathbf{k},\sigma}+\sigma\Omega_{\mathbf{q}}- \xi_{\mathbf{k}+\sigma\mathbf{q},\bar{\sigma}}\right) \,,
\label{eq:tau_up}
\end{align}
where $\int_q = \int \frac{d^3q}{(2\pi)^3}$, $\bar{\sigma} = - \sigma$, $\mathcal{V}$ is the unit cell volume, $n_{F (B)}(\varepsilon)$ the Fermi (Bose) distribution, and $\Omega_\bfq$ is the magnon dispersion (see Supplementary Notes 2 for more details). As mentioned above, for hole-like dispersion this results in $\Gamma_{\text{mag}, \uparrow} > \Gamma_{\text{mag}, \downarrow}$.
The impurity scattering rate
acquires a spin-dependence due to the band splitting included in the renormalized dispersion $\xi_{\bfk, \sigma}$. This makes $\Gamma_{\text{imp}, k_F, \sigma} \propto k_{F, \sigma}$ proportional to the density of states at the Fermi energy~\cite{bruusManyBodyQuantumTheory2004}, and $\Gamma_{\text{imp}, \downarrow}>\Gamma_{\text{imp}, \uparrow}$ for hole-like bands. 

In Fig.~\ref{fig:figure1_theory}(b) and (c), we show results for the total scattering rates $\Gamma_\sigma$, their ratio $\Gamma_\downarrow/\Gamma_{\uparrow}$ and the impurity contribution $\Gamma_{\text{imp}, \sigma}$ at the Fermi energy as a function of $T < T_\text{C}$. We find that for parameters describing EuCd$_2$As$_2$, the ratio is larger than unity, $\Gamma_\downarrow/\Gamma_\uparrow > 1$, (except very close to $T_{\text{C}}$) and approaches two in the low-temperature limit. As shown in Fig.~\ref{fig:figure1_theory}(b), magnetic scattering dominates just below $T_{\text{C}}$, but is suppressed at lower temperatures due to a reduced magnon density. As a result, impurity scattering prevails at low $T$, leading to the observed lifetime enhancement of the majority band $\tau_\uparrow > \tau_{\downarrow}$. 
Both rates are of the order $\Gamma_\sigma \sim 10^{13}$~Hz and increase with temperature, exhibiting a pronounced peak at $T=T_{\text{C}}$ due to strong magnetic scattering. These results are in good agreement with ARPES results and also closely reflect the behavior of the resistivity due to loss of spin disorder scattering (see Fig.~\ref{fig:Fig1}).
We emphasize that the ratio $\Gamma_\downarrow/\Gamma_\uparrow$ increases with decreasing carrier density $n$, making the effect more pronounced in a low density semi-metal such as EuCd$_2$As$_2$.
Finally, taking interband scattering to additional bands into account, our model can also naturally account for the experimentally observed energy dependence of the quasiparticle lifetimes, in particular the fact that the sharp electronic band disappears as soon as a broad additional electronic pocket emerges (see red arrow in Fig.~\ref{fig:Fig2}(a-d)).

\emph{Conclusion.--} We report a direct observation of the evolution of quasiparticle properties into the ferromagnetic phase of the metallic magnet EuCd$_2$As$_2$. In addition to energy shifts and renormalizations proportional to the magnetization, we observe a large enhancement of the carrier's lifetime in selected bands and energy ranges. Investigations of the temperature dependence of the lifetime allow us to quantify different electronic scattering mechanisms and directly observe the impact of magnetic order and magnon excitations on itinerant carriers.
Our work provides a direct link between quasiparticle lifetimes and loss-of-spin-disorder suppression of the resistivity, and reveals a mechanism towards spin selective transport via lifetime tuning of spin polarized carriers.

\section{Data availability}
Relevant data for the work are available at the Materials Data Facility [Ref.]

\section{References and notes}



%

\section{Acknowledgement}

The authors thank R.M. Fernandes, K. O'Neal, and D.A. Yarotski for helpful discussions. This work was supported by the U.S. Department of Energy, Office of Basic Energy Sciences,Division of Materials Sciences and Engineering. This work was also supported by the Center for Advancement of Topological Semimetals, an Energy Frontier Research Center funded by the U.S. Department of Energy Office of Science, Office of Basic Energy Sciences, through the Ames Laboratory under its Contract No. DE-AC02-07CH11358.

\section{Author contributions}
N.H.J and Y.W. contributed equally to the present work. N.H.J. and B.K. grew and characterized the samples under the supervision of S.L.B. and P.C.C.. Y.W., N.H.J., K.L., and B.S. acquired and analyzed ARPES data under the supervision of A.K.. L.-L.W. carried out the DFT calculations. T.V.T and P.P.O. developed the theory modeling and performed analytical calculations. All authors contributed to writing the paper.

\section{Competing financial interests}
The authors declare no competing financial interests.

\section{Additional information}
Correspondence and requests for materials should be addressed to A.K.

\section{Methods}
\textbf{Crystal growth and electrical transport.} Single crystals of EuCd$_{2}$As$_{2}$ were grown using flux growth method as described in Ref.~\onlinecite{Jo2019}. Temperature dependent electrical
transport measurement were carried out in a Quantum Design Physical Property Measurement System (PPMS) for 1.8\,K\,$\leq$ $T$\,$\leq$\,300\,K. The samples for electrical transport measurements were prepared by attaching four Pt wires
using DuPont 4929N silver paint. The current was applied in
ab plane with I = 1 mA and f = 17 Hz.

\textbf{ARPES measurements.}Samples used for ARPES measurements were cleaved \textit{in situ} at 40~K under ultrahigh vacuum (UHV). The data were acquired using a tunable VUV laser ARPES system, that consists of a Omicron Scienta DA30 electron analyzer, a picosecond Ti:Sapphire oscillator and fourth harmonic generator~\cite{Jiang2014Tunable}. Data were collected with photon energies of 6.05 to 6.79~eV. Momentum and energy resolutions were set at $\sim$ 0.005~\AA${}^{-1}$ and 2~meV. The size of the photon beam on the sample was $\sim$30~$\mu$m.

\textbf{DFT calculations.} Band structures with spin-orbit coupling (SOC) in density functional theory (DFT)~\cite{Hohenberg1964Inhomogeneous, Kohn1965Self} have been calculated using a PBE~\cite{Perdew1996Generalized} exchange-correlation functional, a plane-wave basis set and projector augmented wave method~\cite{Blochl1994Projector} as implemented in VASP~\cite{Kresse1996Efficient, Kresse1996Efficiency}. For ferromagnetic (FM) EuCd${}_{2}$As${}_{2}$, a Hubbard-like~\cite{Dudarev1998Electron} U value of 5.0 eV is used to account for the half-filled strongly localized Eu 4$f$ orbitals, while for non-magnetic (NM) EuCd${}_{2}$As${}_{2}$, the Eu 4$f$ orbitals are treated as core electrons. For bulk band structures of a Monkhorst-Pack~\cite{Monkhorst1976Special} ($11\times 11\times 7$) $k$-point mesh with a Gaussian smearing of 0.05~eV including the $\Gamma$ point and a kinetic energy cutoff of 318~eV have been used. Experimental lattice parameters~\cite{Schellenberg2011Mossbauer} have been used with atoms fixed in their bulk positions. A tight-binding model based on maximally localized Wannier functions~\cite{Marzari1997Maximally, Souza2001Maximally, Marzari2012Maximally} was constructed to reproduce closely the bulk band structure including SOC in the range of ${E}_{F}\pm$1~eV with Eu $sdf$, Cd $sp$ and As $p$ orbitals. Then the 2D bulk band dispersions and Fermi surfaces have been calculated with WannierTools~\cite{Wu2017WannierTools}.

\textbf{Theoretical model.}  Our minimal model to describe the observed band-selective sharpening of the ARPES line-width in the ferromagnetic phase of EuCd$_2$As$_2$ is composed by three parts: $H=H_{c}+H_{cf}+H_{f}$. The electronic component,
\begin{equation}
H_c = \sum_{\textbf{k}, \sigma} \xi_\textbf{k} c^\dag_{\textbf{k},\sigma} c_{\textbf{k}, \sigma} + \sum_{\textbf{k}, \textbf{q}, \sigma} \sum_{\textbf{r}_j} v_0 e^{- i (\textbf{k} - \textbf{q}) \cdot \textbf{r}_j} c^\dag_{\textbf{k}, \sigma} c_{\textbf{q}, \sigma} \text{ ,}
\label{eq:Hc}
\end{equation}
\noindent accounts for electrons in a parabolic hole like band, $\xi_\textbf{k} = \hbar^2 k^2/(2 m^*) +W -\mu$, with effective mass $m^*=-m_e$. Here, $m_e$ denotes the rest electron mass and $W$ denotes the energy associated with the top of the band. The electrons experience a random potential $v(\textbf{r}) = v_0 \sum_{j=1}^{N_{\text{imp}}}\delta(\textbf{r} - \textbf{r}_j)$ created by $N_{\text{imp}}$ point-like impurities at random sites $\textbf{r}_j$. Besides,
\begin{equation}
H_{f} = - J_0 \sum_{\av{i,j}} \textbf{S}_i \cdot \textbf{S}_j
\label{eq:Hf}
\end{equation}
\noindent describes $N$ Eu $S=7/2$ magnetic moments on a hexagonal lattice~\cite{Jo2019} with nearest-neighbor ferromagnetic coupling $J_0 > 0$. The interaction between the electrons and the localized moments takes the form
\begin{equation}
H_{cf} = - J \sum_{i} \textbf{S}_i \cdot \textbf{s}_i
\label{eq:Hcf}
\end{equation}
\noindent where $\textbf{s}_i = \frac{\hbar}{2N} \sum_{\textbf{k}, \textbf{k}'} \sum_{\sigma, \sigma'} e^{-i\mathbf{R}_i\cdot (\mathbf{k}-\mathbf{k}')}c^\dag_{\textbf{k}, \sigma} \boldsymbol{\sigma}_{\sigma \sigma'} c_{\textbf{k'}, \sigma'}$ is the spin operator of the conduction electrons, and $\mathbf{R}_i$ denotes the position of site $i$.

At low temperatures, the fluctuations about the ordered phases are small and we can map the previous Hamiltonian into an interacting electron-magnon problem via a Holstein-Primakoff transformation ~\cite{AuerbachInteractingElectronsQuantum1994}, which takes a simpler form $S_{\mathbf{q}}^{z}=\hbar S\sqrt{N}\delta_{\mathbf{q},0}-\hbar\sum_{\mathbf{k}}b_{\mathbf{k}}^{\dag}b_{\mathbf{q}+\mathbf{k}}^{\null}/\sqrt{N}$, $S_{\mathbf{q}}^{+}=S_{\mathbf{q}}^{x}+iS_{\mathbf{q}}^{y}=\hbar \sqrt{2S}b_{\mathbf{q}}$ and $S_{\mathbf{q}}^{-}=S_{\mathbf{q}}^{x}-iS_{\mathbf{q}}^{y}=\hbar \sqrt{2S}b_{-\mathbf{q}}^{\dag}$, when only a small number of magnons are excited in the system. The resulting electrons-magnon interaction has four types of vertices which encode both spin-flip and spin-conserving processes (see Supplementary Information for details) and is treated perturbatively using standard diagrammatic techniques similarly as in Ref.~\onlinecite{woolseyElectronmagnonInteractionFerromagnetic1970}.

Within first-order perturbation theory, we find that the electron band experience a spin-dependent energy split $\xi_\mathbf{k} \rightarrow \xi_{\mathbf{k}, \sigma} = \xi_\mathbf{k} - \sigma \gamma(T)$ due to the effective magnetic field created by the Eu moments in the ordered phase. Since the spin-up band is aligned with the magnetization direction, it is shifted downward and becomes a \textit{majority band}, while the spin-down band is shifted upwards and becomes a \textit{minority band}. The energy shift is temperature-dependent $\gamma(T)= \frac{J\hbar^2}{2} \bigl[S-\frac{\zeta(\frac{3}{2}) (k_BT)^{3/2}}{8 (J_0 \hbar^2S \pi)^{3/2}} \bigr]$ reflecting the fact that as $T$ increases more magnons are excited in the system, which weakens the net magnetization until the ferromagnetic order melts at $T=T_{\text{C}}$ and $\gamma(T_{\text{C}})=0$. These is also a feedback effect of the electron in the magnons, in which the lowest order effect consists of an energy shift of the magnon dispersion proportional to the difference of densities of majority and minority carriers (see Supplementary Notes for more details).

To calculate the magnetic scattering rate we apply second order perturbation theory in the coupling $J$, which results in Eq. (\ref{eq:tau_up}). The integral in Eq.(\ref{eq:tau_up}) was calculated numerically assuming a quadratic dispersion for the magnons $\Omega_{\mathbf{q}}$. The non-magnetic impurities were treated via self-average and the impurity scattering was calculated using the first-order Born approximation, yielding (see Supplementary Information for details)
\begin{equation}
\Gamma_{\text{imp}, \mathbf{k}, \sigma} = \frac{\pi n_{\text{imp}} |v_0|^2}{\hbar}\frac{1}{V} \sum_{\mathbf{q}} \delta(\xi_{\mathbf{q}, \sigma} - \xi_{\mathbf{k}, \sigma}) \text{ ,}
\end{equation}
\noindent where $n_{imp}=N_{imp}/V$ denotes the impurity density, where $V$ is the system volume. We set $n_{imp}|v_0|^2\approx 2\times 10^{-24}$ eV$^2$cm$^3$ in Fig.~\ref{fig:figure1_theory}, which we estimated from the experimental quasiparticle scattering rate $\approx 0.08$ ps for the majority band.

\newpage

{\bf Supplementary Information for ``Visualizing band selective enhancement of quasiparticle lifetime in a metallic ferromagnet'}

\newpage

\section{Supplementary Note 1. Interacting electron-magnon model}

In this section, we derive the Hamiltonian of the interacting electron-magnon problem, which is valid at \textit{low} temperatures ($T<T_{\text{C}}$), when the fluctuations around the ordered phase are not too strong. In this regime, the spin operators $S_i^{\alpha}$ (with $\alpha=x,y,z$) of the localized Eu moments can be mapped, via a Holstein-Primakoff transformation, into bosonic (magnons) creation and annihilation operators that describe excitation about the ordered magnetic phase,
\begin{align}
&S_{j}^{z}=\hbar(S-\hat{n}_{j})\text{ ,}\label{eq:HPSz}\\
&S_{j}^{+}=S_{j}^{x}+iS_{j}^{y}=\hbar\sqrt{2S-\hat{n}_j}b_j \text{ ,}\\
&S_{j}^{-}=S_{j^{x}}-iS_{j}^{y}=\hbar b_j^{\dag}\sqrt{2S-\hat{n}_j}\text{ .}\label{eq:HPS-}
\end{align}

\noindent Here, $\hat{n}_j=b_{j}^{\dag}b_{j}^{\null}$ is the bosonic number operator. These expressions can be further simplified by the approximation $\sqrt{2S-\hat{n}_i}\approx \sqrt{2S}$, valid when the number of bosons excited in the system is not very large. In this case, in momentum space, we find
\begin{align}
& S_{\mathbf{q}}^{z}=\hbar S\sqrt{N}\delta_{\mathbf{q},0}-\frac{\hbar}{\sqrt{N}}\sum_{\mathbf{k}}b_{\mathbf{k}}^{\dag}b_{\mathbf{q}+\mathbf{k}}^{\null} \text{ ,}\label{eq:Sz_momentum}\\
& S_{\mathbf{q}}^{+}=\hbar \sqrt{2S}b_{\mathbf{q}} \text{ ,}\label{eq:SP_momentum}\\
& S_{\mathbf{q}}^{-}=\hbar \sqrt{2S}b_{-\mathbf{q}}^{\dag} \label{eq:SM_momentum}\text{ ,}
\end{align}

\noindent where
\begin{align}
& b_{i}^{\dag}=\frac{1}{\sqrt{N}}\sum\limits_{\mathbf{k}}e^{-i\mathbf{k}\cdot \mathbf{R}_i}b_{\mathbf{k}}^{\dag} \text{ ,} \\
& S_{\mathbf{q}}^{\alpha}=\frac{1}{\sqrt{N}}\sum\limits_{i=1}^{N}e^{-i\mathbf{q}\cdot\mathbf{R}_{i}}S_{i}^{\alpha} \text{ .}
\end{align}

\noindent Recall that $\mathbf{R}_i$ denotes the position of the $N$ localized Eu moments in an hexagonal lattice.

\begin{figure}[t!]
\centering
\includegraphics[width=0.2\textwidth]{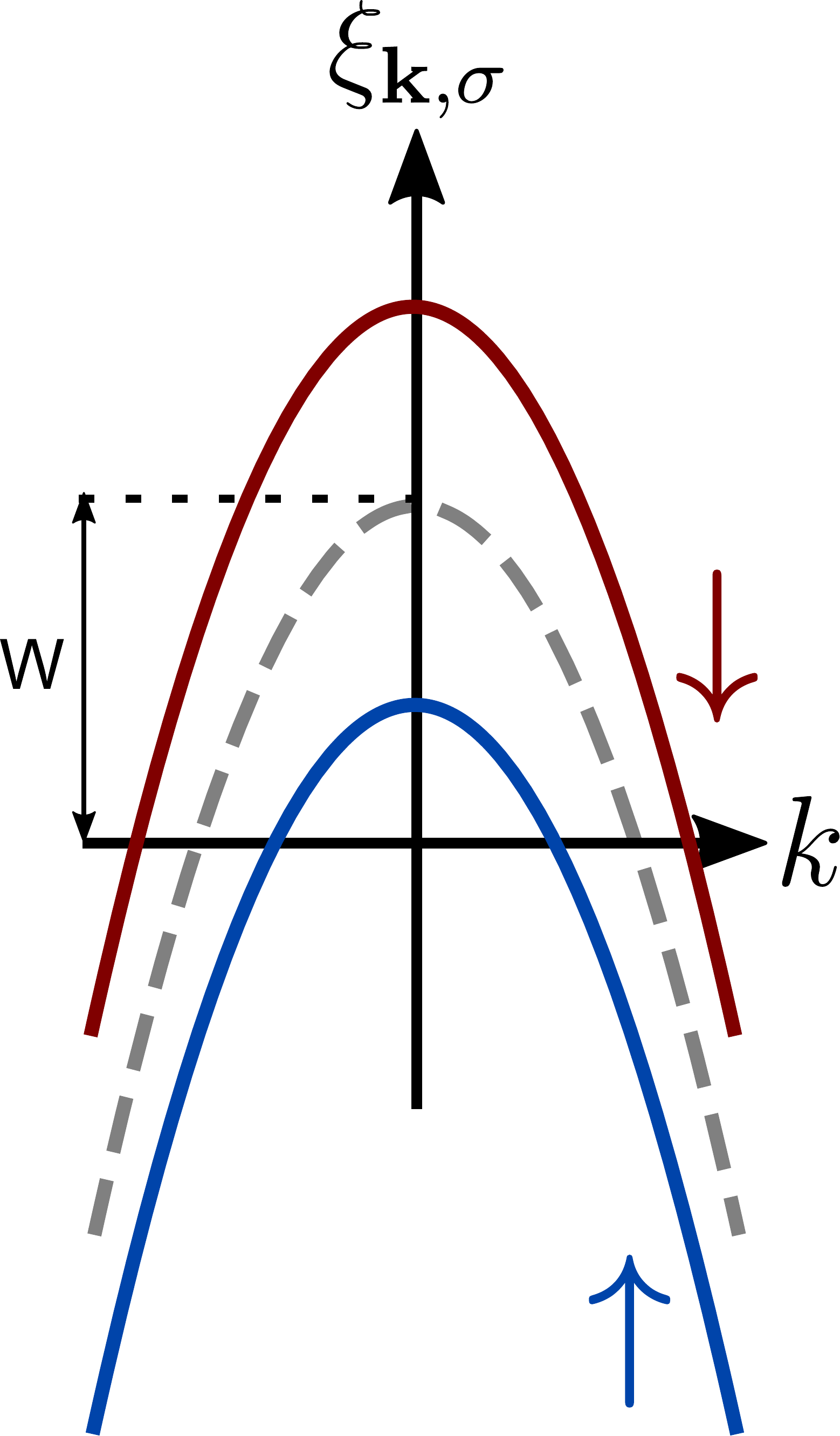}
\caption{Electron dispersion of our effective model. The dashed line correspond to a spin-degenerate parabolic hole-like band (effective mass $m^*<0$). $W$ corresponds to the energy of the top of the band. The solid lines correspond the spin split of the band in majority (spin up) and minority (spin down) bands, which occur in the ferromagnetic phase, as discussed in the Supplementary Note 3.}
\label{fig:bands}
\end{figure}

Substituting Eqs.(\ref{eq:Sz_momentum})-(\ref{eq:SM_momentum}) into Eqs.(2)-(4) of the Methods Section we get\footnote{Because of the periodicity of the magnetic lattice $\sum\limits_{i=1}^{N}e^{-i(\mathbf{k}-\mathbf{q})\cdot \mathbf{R}_i}=N\delta_{\mathbf{k},\mathbf{q}}$.}:
\begin{align}
&H_c = \sum_{\textbf{k}, \sigma} \xi_\textbf{k} c^\dag_{\textbf{k},\sigma} c_{\textbf{k}, \sigma} + \sum_{\textbf{k}, \textbf{q}, \sigma} \sum_{\textbf{r}_j} v_0 e^{- i (\textbf{k} - \textbf{q}) \cdot \textbf{r}_j} c^\dag_{\textbf{k}, \sigma} c_{\textbf{q}, \sigma} \text{ ,}\label{eq:HcSI}\\
&H_{f}=-\frac{J_0\hbar^2S^2Nz}{2}+\sum\limits_{\mathbf{q}}\Omega^{(0)}_{\mathbf{q}}\,b_{\mathbf{q}}^{\dag}b_{\mathbf{q}}^{\null} \text{ ,}
\label{eq:Heisenberg2}\\
&H_{cf}=\sum\limits_{\alpha,\beta}\sum\limits_{\mathbf{k},\mathbf{q}}\Gamma_{\alpha,\beta}(\mathbf{k},\mathbf{q})c_{\mathbf{k}\alpha}^{\dag}c_{\mathbf{q}\beta}^{\null} \text{ .}
\label{eq:Hint_momentum}
\end{align}

\noindent Here, $\xi_\textbf{k} = \hbar^2 k^2/(2 m^*) +W -\mu$ is the hole-like parabolic electron dispersion, as illustrated in Fig.\ref{fig:bands}. We denote by $m^*=-m_e$ the effective electron mass, $m_e$ is the electron rest mass, $\mu$ is the chemical potential that controls the occupation of the band, and $W$ denotes the energy of the top of the band. Besides,
\begin{equation}
\Omega_{\mathbf{q}}^{(0)}=J_0\hbar^2S\left(z-2\sum\limits_{\boldsymbol{\delta}}\cos(\mathbf{q}\cdot\boldsymbol{\delta})\right) \text{ ,}\label{eq:Omega0a}
\end{equation}

\noindent is the magnon dispersion, which will be latter approximated by a quadratic dispersion (see Supplementary Note $4$), and the vector $\boldsymbol{\delta}$ connects nearest neighbor sites of the magnetic lattice. Higher-order terms in $\hat{n}_i$ gives rise to magnon-magnon interaction, which we do not take into account in this model.

The vertex of the electron-magnon interaction $\Gamma_{\alpha,\beta}(\mathbf{k},\mathbf{q})$ is the combination of four processes,
\begin{equation}
\Gamma_{\alpha,\beta}(\mathbf{k},\mathbf{q})=\Gamma^{(z)}_{\alpha,\beta}(\mathbf{k},\mathbf{q})+\Gamma^{(+)}_{\alpha,\beta}(\mathbf{k},\mathbf{q})+\Gamma^{(-)}_{\alpha,\beta}(\mathbf{k},\mathbf{q})+\tilde{\Gamma}_{\alpha,\beta}(\mathbf{k},\mathbf{q}) \text{ ,}
\end{equation}

\noindent with
\begin{align}
& \Gamma^{(z)}_{\alpha,\beta}(\mathbf{k},\mathbf{q})\equiv -\frac{J\hbar^2}{2}S\,\sigma_{\alpha,\beta}^{z}\,\delta_{\mathbf{k},\mathbf{q}} \text{ ,}\label{eq:Gammaz}\\[0.2cm]
& \Gamma^{(+)}_{\alpha,\beta}(\mathbf{k},\mathbf{q})\equiv -\frac{J\hbar^2}{2}\sqrt{\frac{S}{2N}}\sigma_{\alpha,\beta}^{+}\, b_{\mathbf{q}-\mathbf{k}}^{\dag} \text{ ,}\label{eq:GammaPlus}\\[0.2cm]
& \Gamma^{(-)}_{\alpha,\beta}(\mathbf{k},\mathbf{q})\equiv -\frac{J\hbar^2}{2}\sqrt{\frac{S}{2N}}\sigma_{\alpha,\beta}^{-}\, b_{\mathbf{k}-\mathbf{q}}^{\null} \text{ ,}\label{eq:GammaMinus}\\[0.2cm]
& \tilde{\Gamma}_{\alpha,\beta}(\mathbf{k},\mathbf{q})\equiv\frac{J\hbar^2}{2N}\sigma_{\alpha,\beta}^{z}\sum\limits_{\mathbf{k}'}b_{\mathbf{k}'}^{\dag}b_{\mathbf{k}'+\mathbf{k}-\mathbf{q}}^{\null} \text{ .}\label{eq:tildeGamma}
\end{align}

\begin{figure}[b!]
\centering
\includegraphics[width=0.99\textwidth]{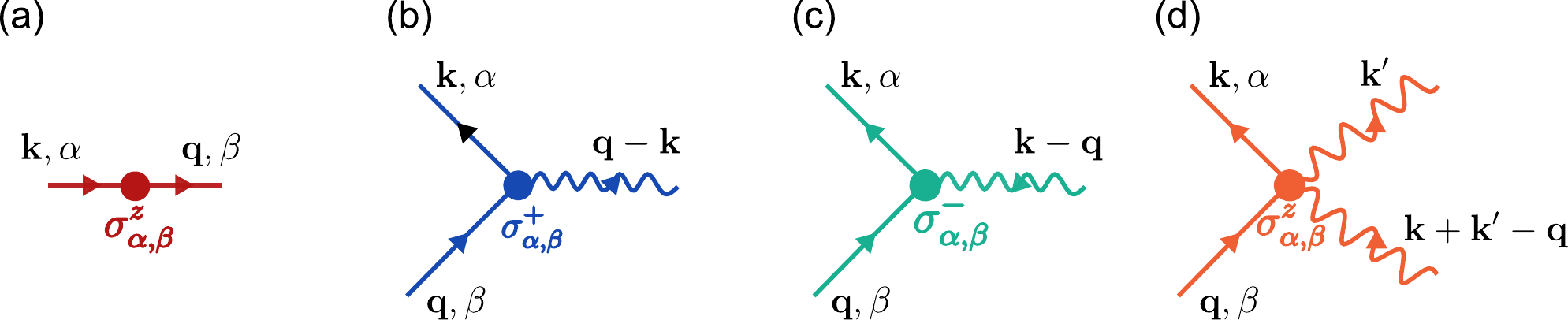}
\caption{Vertex of the electron-magnon interaction defined in Eqs.~(\ref{eq:Gammaz})-(\ref{eq:tildeGamma}).}
\label{fig:vertices}
\end{figure}

\noindent These processes are illustrated in Fig.\ref{fig:vertices}. Note that $\Gamma^{(z)}$  and $\tilde{\Gamma}$ are spin-conserving processes, while $\Gamma^{(+)}$ ($\Gamma^{(-)}$) involves the creation (absorption) of a magnon and require a flip of the electron spin. In the previous equations, $\sigma^{\pm}=\sigma^{x}\pm i\sigma^{y}$, where $\sigma^{x}$, $\sigma^{y}$ and $\sigma^{z}$ are the Pauli matrices.

\section{Supplementary Note 2. Perturbation theory}\label{sec:pt}
In this section we treat $H_{cf}$ using standard diagrammatic techniques to investigate the effects of magnons on the electronic degrees of freedom up to the order $J^2$. We also address the feedback effect of the electrons in the magnon propagator up to linear order in $J$. The calculations performed in this section follow those of \textit{Woolsey et al. PRB \textbf{1}, 11 (1970)}.

\subsection{Bosonic degrees of freedom}

\begin{figure}[b!]
\centering
\includegraphics[width=0.8\textwidth]{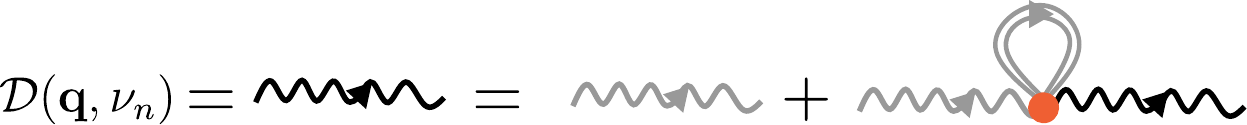}
\caption{Dyson equation for the magnon propagator. The interaction vertex follow the same color code as in Fig.\ref{fig:vertices}. The gray wiggle line correspond to the bare magnon propagator $\mathcal{D}^{(0)}(\mathbf{q},\nu_n)$ defined in Eq.(\ref{eq:D0}), while the black wiggle line represent the dressed bosonic propagator up to first order in $J$. The solid gray double line correspond to the first-order fermion Green's function (see next section).}
\label{fig:Dyson_magnon}
\end{figure}

The (finite-temperature $T$) propagator of free magnons is given by
\begin{equation}
\mathcal{D}^{(0)}(\mathbf{q},\nu_n)=\frac{1}{i\nu_{n}-\Omega^{(0)}_{\mathbf{q}}/\hbar} \text{ ,}
\label{eq:D0}
\end{equation}

\noindent where $\nu_n=2n\pi k_BT/\hbar$ (integer $n$) are the bosonic Matsubara frequencies and $\Omega^{(0)}(\mathbf{q})$ is the free magnon dispersion defined in Eq.(\ref{eq:Omega0a}). In lowest-order perturbation theory (RPA-like Dyson equation shown in Fig.~\ref{fig:Dyson_magnon}), the electron-magnon interaction promotes a energy shift in the magnon dispersion: $\Omega(\mathbf{q})=\omega_0+\Omega^{(0)}(\mathbf{q})$, with
\begin{equation}
\omega_0=J\hbar^2\frac{\mathcal{V}}{2}(n_{\uparrow}-n_{\downarrow}) \text{ .}
\end{equation}

\noindent Here, $n_{\uparrow}$ ($n_{\downarrow}$) denotes the density of electrons with spin projection up (down). Besides, $\mathcal{V}=V/N$ is the volume of the direct magnetic cell and $V$ is the total volume of the system.

\subsection{Electronic degrees of freedom}

\begin{figure}[b!]
\centering
\includegraphics[width=0.85\textwidth]{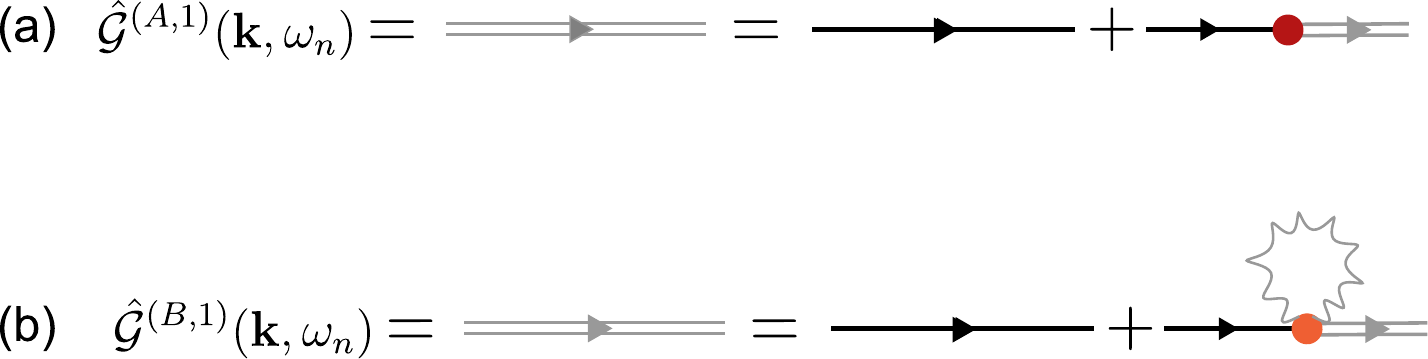}
\caption{Dyson equations (\ref{eq:G1}) and (\ref{eq:G1B}) for the electron Green's function up to first order in $J$. $\hat{\mathcal{G}}$ denotes a matrix in spin-space whose matrix elements are $\mathcal{G}_{\alpha,\beta}$ specified in the text. The interaction vertex follow the same color code as in Fig.\ref{fig:vertices}. The gray wiggle line correspond to the bare magnon propagator defined in Eq.(\ref{eq:D0}), while the black solid line represent bare fermion Green's function defined in Eq.(\ref{eq:G0}).}
\label{fig:Dyson_electron_first}
\end{figure}

To investigate the effects of magnons on the electrons, we start by calculating the electronic Green's function dressed by self-energy insertions that are built using only the vertex $\Gamma^{(z)}$, as shown in Fig.\ref{fig:Dyson_electron_first}(a),
\begin{equation}
\mathcal{G}^{(A,1)}_{\alpha,\beta}(\mathbf{k},\omega_n)=\left[\mathcal{G}^{(0)\,-1}_{\alpha,\beta}(\mathbf{k},\omega_n)-\Sigma_{\alpha,\beta}^{(A,1)}(\mathbf{k},\omega_n)\right]^{-1} \text{ ,}
\label{eq:G1}
\end{equation}

\noindent where
\begin{equation}
\mathcal{G}^{(0)}_{\alpha,\beta}(\mathbf{k},\omega_n)=\frac{\delta_{\alpha,\beta}}{i\omega_n-\xi_{\mathbf{k}}/\hbar}
\label{eq:G0}
\end{equation}

\noindent is the bare electron Green's function, with $\omega_{n}=(2n+1)\pi k_BT/\hbar$ ($n$ integer) denoting the fermionic Matsubara frequency, and
\begin{equation}
\Sigma_{\alpha,\beta}^{(A,1)}(\mathbf{k},\omega_n)=-\frac{\gamma_0}{\hbar}\sigma_{\alpha,\beta}^{(z)} \text{ ,}
\label{eq:S1}
\end{equation}

\noindent where we define $\gamma_0\equiv J\hbar^2S/2$. The Dyson's equation (\ref{eq:G1}) gives us the electron Green's function in first-order perturbation theory,
\begin{align}
&\mathcal{G}^{(A,1)}_{\uparrow,\uparrow}(\mathbf{k},\omega_n)=\frac{1}{i\omega_n-(\xi_{\mathbf{k}}-\gamma_0)/\hbar}\text{ ,}\label{eq:G1upup}\\[0.2cm]
&\mathcal{G}^{(A,1)}_{\downarrow,\downarrow}(\mathbf{k},\omega_n)=\frac{1}{i\omega_n-(\xi_{\mathbf{k}}+\gamma_0)/\hbar}\text{ ,}\\[0.2cm]
&\mathcal{G}^{(A,1)}_{\uparrow,\downarrow}(\mathbf{k},\omega_n)=\mathcal{G}^{(0)}_{\downarrow,\uparrow}(\mathbf{k},\omega_n)=0 \text{ ,} \label{eq:G1updown}
\end{align}

\noindent which tell us that the effect of the electron-magnon interaction is to generate a Zeeman-like energy shift of the electron bands $\xi_{\mathbf{k}}\rightarrow\xi_{\mathbf{k},\sigma}=\xi_{\mathbf{k}}-\gamma_0\sigma$, where we identify $\uparrow=+$ and $\downarrow=-$. As discussed in the main text, this causes the spin-up band to shift down in energy and it becomes a \textit{majority band}, since it can now accommodate more electrons. The spin-down band, on the other hand, shifts up in energy and becomes and corresponds to a \textit{minority band}.

This result can be further improved. The shift of the electron bands cannot be really a constant, but it is rather a temperature-dependent function. The reasoning is the following: as the temperature increases, more spin-flip processes take place, causing ground-state magnetization to decrease. Equivalently, more magnons are introduced in the system as temperature increases, until the ferromagnetic order melts at the transition temperature $T=T_{\text{C}}$. As a consequence, the effective magnetic field felt by the electrons decreases with temperature, and so does the band-splitting. This effect is captured by another set of first-order diagrams involving the vertex $\tilde{\Gamma}_{\alpha,\beta}(\mathbf{k},\mathbf{q})$ defined in Eq.(\ref{eq:tildeGamma}), as illustrated in Fig.\ref{fig:Dyson_electron_first} (b). The Dyson equation we have to solve, in this case, is
\begin{align}
&\mathcal{G}^{(B,1)}_{\alpha,\beta}(\mathbf{k},\omega_n)=\left[\mathcal{G}^{(0)\,-1}_{\alpha,\beta}(\mathbf{k},\omega_n)-\Sigma_{\alpha,\beta}^{(B,1)}(\mathbf{k},\omega_n)\right]^{-1} \text{ ,}\label{eq:G1B}\\[0.2cm]
&\Sigma_{\alpha,\beta}^{(B,1)}(\mathbf{k},\omega_n)=-\frac{k_BT}{\hbar^2}\frac{J\hbar^2}{2N}V\sigma_{\alpha,\beta}^{(z)}\sum\limits_{n'}\int\frac{d^3q}{(2\pi)^3}\mathcal{D}^{(0)}(\mathbf{q},\nu_n') \text{ ,}\label{eq:S1B}
\end{align}

\noindent where $\mathcal{D}^{(0)}(\mathbf{q},\nu_n)$ is the bosonic propagator defined in Eq.(\ref{eq:D0}).

The Matubara sum in Eq.(\ref{eq:S1B}) can be easily calculated:
\begin{equation}
\frac{k_B T}{\hbar}\sum\limits_{n} \frac{1}{i\nu_n-\Omega^{(0)}_{\mathbf{q}}/\hbar}=\int_{\mathcal{C}}\frac{dz}{2\pi i}\frac{1}{e^{z/(k_B T)}-1}\frac{1}{z-\Omega^{(0)}_{\mathbf{q}}}=-n_B(\Omega^{(0)}_{\mathbf{q}}) \text{ ,}
\end{equation}

\noindent where $n_{B}(\epsilon)=\left(e^{\epsilon/(k_B T)}-1\right)^{-1}$ is the Bose-Einstein distribution function. The remaining momentum integration can also be evaluated analytically when the boson dispersion is approximated to a parabola~\cite{auerbachInteractingElectronsQuantum1994},
\begin{equation}
\frac{V}{N}\int \frac{d^3q}{(2\pi)^3} n_B(\Omega^{(0)}_\mathbf{q})=\frac{1}{8}\left(\frac{k_BT}{J_0\hbar^2S\pi}\right)^{3/2}\zeta(3/2) \text{ .}
\end{equation}

\noindent Therefore,
\begin{equation}
\Sigma_{\alpha,\beta}^{(B,1)}(\mathbf{k},\omega_n)=\frac{1}{\hbar}\frac{J\hbar^2}{2}\sigma_{\alpha,\beta}^{(z)}\frac{1}{8}\left(\frac{k_B T}{J_0\hbar^2S\pi}\right)^{3/2}\zeta(3/2) \text{ ,}
\label{eq:S1B}
\end{equation}

\noindent where $\zeta(x)$ is the Riemann zeta function.

Combining Eqs.(\ref{eq:S1}) and (\ref{eq:S1B}), we obtain
\begin{align}
& \Sigma_{\alpha,\beta}^{(1)}(\mathbf{k},\omega_n)=-\frac{1}{\hbar}\frac{J\hbar^2}{2}\left\langle S_z\right\rangle \text{ ,}\\
& \left\langle S_z\right\rangle=S-\frac{1}{8}\left(\frac{k_B T}{J_0S\hbar^2\pi}\right)^{3/2}\zeta(3/2) \,.
\end{align}

\noindent The corresponding dressed first-order Green's function has the same form of Eqs.(\ref{eq:G1upup})-(\ref{eq:G1updown}), but with a $T$-dependent shift of the bands $\xi_{\mathbf{k},\sigma}=\xi_{\mathbf{k}}-\gamma(T)\sigma$, with
\begin{equation}
\gamma(T)\equiv \frac{J\hbar^2}{2}\left(S-\frac{1}{8}\left(\frac{k_B T}{J_0\hbar^2S\pi}\right)^{3/2}\zeta(3/2)\right)  \text{ .}
\label{eq:xi_sigma}
\end{equation}

\noindent Importantly, when $T=T_{\text{C}}$, $\gamma(T_{\text{C}})=0$, which allow us to identify
\begin{equation}
J_0 \hbar^2S=\frac{k_B T_{\text{C}}}{\pi}\left(\frac{\zeta(3/2)}{8S}\right)^{2/3} \text{ .}
\end{equation}

\noindent For EuCd$_2$As$_2$, $T_{\text{C}}\approx 26K$ and $S=7/2$, therefore $J_0S\hbar^2\approx 0.15 meV$.

\begin{figure}[b!]
\centering
\includegraphics[width=0.85\textwidth]{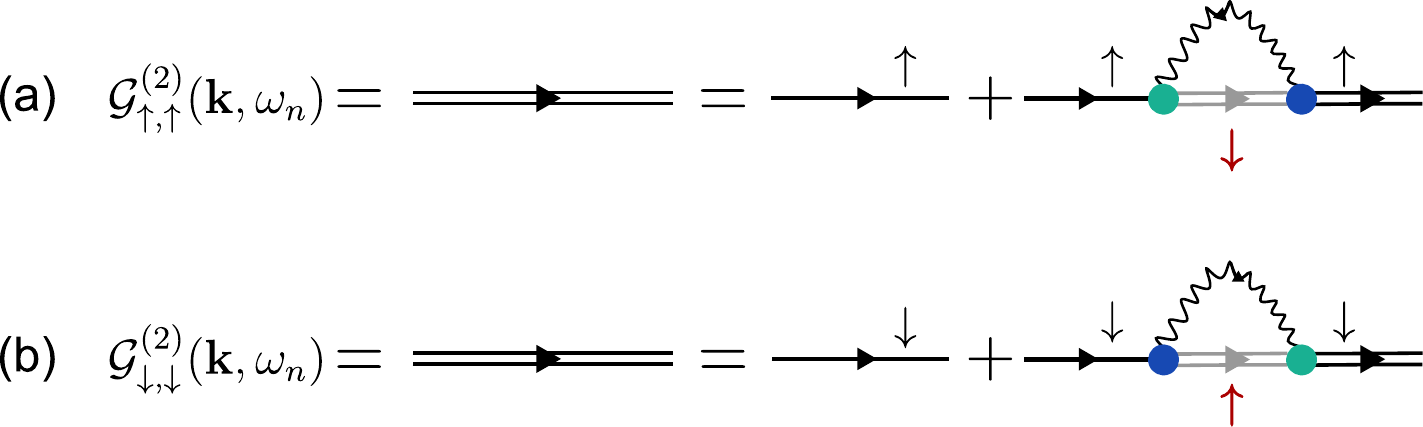}
\caption{Dyson equations (\ref{eq:Dyson2}) for the electron Green's function up to second order in $J$. The interaction vertex follow the same color code as in Fig.\ref{fig:vertices}. The black wiggle line correspond to the dressed magnon propagator and the single solid line correspond to the bare electron Green's function defined in Eq.(\ref{eq:G0}). The double gray (black) solid line, on the other hand, correspond to the electron Green's function up to first (second) order in $J$.}
\label{fig:Dyson_electron_2nd}
\end{figure}

We now focus on second-order perturbation theory, which, as we will shortly show, leads to a spin-dependent scattering of the electrons due to magnons, which we call \textit{magnetic scattering rate}. Diagrams involving the $\Gamma^{(+)}_{\alpha,\beta}(\mathbf{k},\mathbf{q})$ and $\Gamma^{(-)}_{\alpha,\beta}(\mathbf{k},\mathbf{q})$ [see Eqs.(\ref{eq:GammaPlus}) and (\ref{eq:GammaMinus})] requires at least two vertex, so the magnon absorption and emission lines can be combined. Below, we consider only the rainbow diagrams shown in Fig.\ref{fig:Dyson_electron_2nd}. The second-order electronic Green's function is then given by the Dyson equation
\begin{equation}
\mathcal{G}^{(2)}_{\alpha,\beta}(\mathbf{k},\omega_n)=\left[\mathcal{G}^{(0)\,-1}_{\alpha,\beta}(\mathbf{k},\omega_n)-\Sigma_{\alpha,\beta}^{(2)}(\mathbf{k},\omega_n)\right]^{-1} \text{ ,}
\label{eq:Dyson2}
\end{equation}

\noindent with
\begin{align}
&\Sigma_{\uparrow,\uparrow}^{(2)}(\mathbf{k},\omega_n)=-\frac{k_BT}{\hbar^3}V\sum\limits_{n'}\sum\limits_{\alpha',\beta'}\int\frac{d^3q}{(2\pi)^3}\mathcal{D}(\mathbf{q}-\mathbf{k},\omega_n'-\omega_n)\Gamma^{(+)}_{\alpha,\alpha'}(\mathbf{k},\mathbf{q})\mathcal{G}^{(1)}_{\alpha'\beta'}(\mathbf{q},\omega_n')\Gamma^{(-)}_{\beta',\beta}(\mathbf{q},\mathbf{k}) \text{ ,}\label{eq:S2up}\\[0.2cm]
&\Sigma_{\downarrow,\downarrow}^{(2)}(\mathbf{k},\omega_n)=-\frac{k_BT}{\hbar^3}V\sum\limits_{n'}\sum\limits_{\alpha',\beta'}\int\frac{d^3q}{(2\pi)^3}\mathcal{D}(\mathbf{k}-\mathbf{q},\omega_n-\omega_n')\Gamma^{(-)}_{\alpha,\alpha'}(\mathbf{k},\mathbf{q})\mathcal{G}^{(1)}_{\alpha'\beta'}(\mathbf{q},\omega_n')\Gamma^{(+)}_{\beta',\beta}(\mathbf{q},\mathbf{k}) \text{ .}\label{eq:S2down}
\end{align}

\noindent Here, $\mathcal{D}(\mathbf{q}-\mathbf{k},\omega_n'-\omega_n)$ is the magnon propagator with the same form of Eq.(\ref{eq:D0}), but with the remormalized dispersion $\Omega_{\mathbf{q}}=\omega_0+\Omega_{\mathbf{q}}^{(0)}$.

Substituting Eqs.(\ref{eq:GammaPlus}), (\ref{eq:GammaMinus}), (\ref{eq:D0}) and (\ref{eq:G0}) into Eqs.(\ref{eq:S2up})-(\ref{eq:S2down}), we obtain, after evaluating the Matsubara sum,
\begin{align}
& \Sigma_{\uparrow,\uparrow}^{(2)}(\mathbf{k},\omega_n)=V\frac{J^2\hbar^4S}{2N}\frac{1}{\hbar^2}\int \frac{d^3q}{(2\pi)^3}\frac{n_F(\xi_{\mathbf{q},\downarrow})+n_{B}(\Omega_{\mathbf{q}-\mathbf{k}})}{i\omega_n-(\xi_{\mathbf{q},\downarrow}/\hbar)+(\Omega_{\mathbf{q}-\mathbf{k}}/\hbar)}\\[0.2cm]
& \Sigma_{\downarrow,\downarrow}^{(2)}(\mathbf{k},\omega_n)=V\frac{J^2\hbar^4S}{2N}\frac{1}{\hbar^2}\int \frac{d^3q}{(2\pi)^3}\frac{1+n_{B}(\Omega_{\mathbf{k}-\mathbf{q}})-n_F(\xi_{\mathbf{q},\uparrow})}{i\omega_n-(\xi_{\mathbf{q},\uparrow}/\hbar)-(\Omega_{\mathbf{k}-\mathbf{q}}/\hbar)}\text{ .}
\end{align}

\noindent From these finite-$T$ self-energies, we can obtain the magnetic scattering rates after performing the analytic continuation $i\omega_n\rightarrow \omega+i\text{sgn}(\omega_n)\eta$, with $\eta\rightarrow 0^+$, and using the definition of the quasi-particle scattering rate~\cite{bruusManyBodyQuantumTheory2004}
\begin{equation}
\frac{1}{2\tau_{\mathbf{k}}}=-\text{sgn}(\omega_n)\left.\text{Im}\Sigma(\mathbf{k},\omega+i\text{sgn}(\omega_n)\eta)\right|_{\omega=\xi_{k}/\hbar} \text{ .}
\end{equation}

\noindent The fact that we have different self-energies for spin-up and spin-down electrons reflect in spin-dependent magnetic scattering rates:
\begin{align}
&\Gamma_{mag,\mathbf{k},\uparrow}\equiv \frac{1}{2\tau_{mag,\mathbf{k},\uparrow}}=\frac{\pi J^2\hbar^4S\mathcal{V}}{2\hbar}\int \frac{d^3q}{(2\pi)^3}\left[n_F(\xi_{\mathbf{q},\downarrow})+n_{B}(\Omega_{\mathbf{q}-\mathbf{k}})\right]\delta\left(\xi_{\mathbf{k},\uparrow}+\Omega_{\mathbf{q}-\mathbf{k}}-\xi_{\mathbf{q},\downarrow}\right)\label{eq:tau_up_v1}\\[0.2cm]
&\Gamma_{mag,\mathbf{k},\downarrow}\equiv  \frac{1}{2\tau_{mag,\mathbf{k},\downarrow}}=\frac{\pi J^2\hbar^4S\mathcal{V}}{2\hbar}\int \frac{d^3q}{(2\pi)^3}\left[1+n_{B}(\Omega_{\mathbf{k}-\mathbf{q}})-n_F(\xi_{\mathbf{q},\uparrow})\right]\delta\left(\xi_{\mathbf{k},\downarrow}-\Omega_{\mathbf{k}-\mathbf{q}}-\xi_{\mathbf{q},\uparrow}\right) \text{ ,}\label{eq:tau_down_v1}
\end{align}

\noindent Making the change of variables $\mathbf{q}-\mathbf{k}\rightarrow \mathbf{q}$ and $\mathbf{k}- \mathbf{q}\rightarrow  \mathbf{q}$ in Eq.(\ref{eq:tau_up_v1}) and Eq.(\ref{eq:tau_down_v1}), respectively, we find
\begin{align}
& \Gamma_{mag,\mathbf{k},\uparrow}=\frac{\pi J^2\hbar^4S\mathcal{V}}{2\hbar}\int \frac{d^3q}{(2\pi)^3}\left[n_F(\xi_{\mathbf{q}+\mathbf{k},\downarrow})+n_{B}(\Omega_{\mathbf{q}})\right]\delta\left(\xi_{\mathbf{k},\uparrow}+\Omega_{\mathbf{q}}-\xi_{\mathbf{q}+\mathbf{k},\downarrow}\right)\label{eq:tau_up_v2}\\[0.2cm]
&\Gamma_{mag,\mathbf{k},\downarrow}=\frac{\pi J^2\hbar^4S\mathcal{V}}{2\hbar}\int \frac{d^3q}{(2\pi)^3}\left[1+n_{B}(\Omega_{\mathbf{q}})-n_F(\xi_{\mathbf{k}-\mathbf{q},\uparrow})\right]\delta\left(\xi_{\mathbf{k},\downarrow}-\Omega_{\mathbf{q}}-\xi_{\mathbf{k}-\mathbf{q},\uparrow}\right) \text{ ,}\label{eq:tau_down_v2}
\end{align}

\noindent For the results shown in the main text, we evaluate the integrals in Eqs.(\ref{eq:tau_up_v2})-(\ref{eq:tau_down_v2}) at the Fermi level assuming a quadratic dispersion for the bosons,
\begin{equation}
\Omega_{\mathbf{q}}=\omega_0+\frac{\hbar^2 q^2}{2M}\text{ ,}
\label{eq:quadratic_bosons}
\end{equation}

\noindent where $M$ denotes the effective magnon mass [see Supplementary Note 4]. In this case, there is a natural upper cutoff for the momentum integral, $0\leq q\leq q_{max}$, where $4\pi q_{max}^3/3=V_{1BZ}$ and $V_{1BZ}$ denotes the volume of the first Brillouin zone of the magnetic lattice, above which the parabolic approximation breaks down.

Note that the magnetic scattering rate for spin-up electrons depend on the phase space available for scattering to intermediary spin-down states and vice versa. Since, for hole-like bands, the Fermi surface of spin-down electrons (minority) is larger than that of spin-up spins (majority), it follows that $\Gamma_{mag,\uparrow}>\Gamma_{mag,\downarrow}$. The result would be the opposite for electron-like bands ($m^{*}>0$).

\section{Supplementary Note 3. Impurity scattering}
\begin{figure}[b!]
\centering
\includegraphics[width=0.8\textwidth]{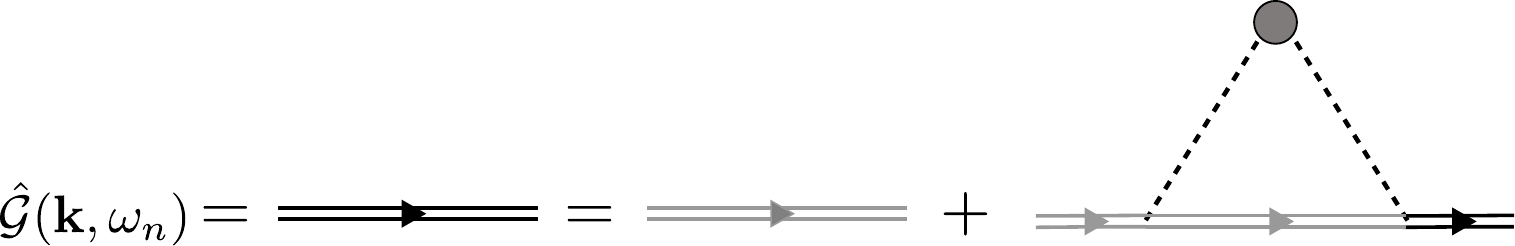}
\caption{Dyson equations (\ref{eq:Dyson_disorder}) for the electron Green's function dressed by disorder according to the first-order Born approximation. The gray double solid line correspond to the fist-order Green's function $\hat{\mathcal{G}}^{A,1}(\mathbf{k},\omega_n)+\hat{\mathcal{G}}^{B,1}(\mathbf{k},\omega_n)$. The dotted line represent the impurity potential $v_0$ and the solid gray circle signifies the impurity density $n_{imp}$.}
\label{fig:Dyson_disorder}
\end{figure}

Another channel for electron scattering in EuCd$_2$As$_2$ is the impurity scattering. To calculate the \textit{impurity scattering rate}, we consider a total of $N_{imp}$ non-magnetic impurities placed at random positions $\boldsymbol{\tau}_j$. For simplicity, we focus on the case of point-like disorder, so the impurity potential felt by the electrons takes the form of the second term in Eq.(\ref{eq:HcSI}). Disorder then dresses the electron Green's function according to
\begin{equation}
\mathcal{G}_{\alpha,\beta}(\mathbf{k},\omega_n)=\left[\mathcal{G}^{(1)\,-1}_{\alpha,\beta}(\mathbf{k},\omega_n)-\Sigma_{\alpha,\beta}^{(imp)}(\mathbf{k},\omega_n)\right]^{-1} \text{ ,}
\label{eq:Dyson_disorder}
\end{equation}

\noindent where the impurity self-energy is obtained via standard Born-approximation, as illustrated in Fig.\ref{fig:Dyson_magnon}. Here, $\mathcal{G}^{(1)}_{\alpha,\beta}(\mathbf{k},\omega_n)$ is the first-order Green's function calculated in Supplementary Note 2. Because of the spin-split of the band in the ferromagnetic phase, impurities scatter spin-up electrons and spin-down electrons differently. As we will shortly see, as in the case of the magnetic scattering rate, this is related to different phase spaces of spin-up and spin-down electron.

The impurity self-energy is given by
\begin{equation}
\Sigma_{\alpha,\beta}^{(imp)}(\mathbf{k},\omega_n)=\delta_{\alpha,\beta}\frac{n_{imp}|v_0|^2}{\hbar}\frac{1}{V}\sum\limits_{\mathbf{k}'}\frac{1}{i\omega_n-\xi_{\mathbf{k}',\alpha}} \text{ .}
\end{equation}

\noindent Performing the analytic continuation as in Supplementary Note 2, we find
\begin{equation}
\Gamma_{imp,\mathbf{k},\sigma}\equiv\frac{1}{2\tau_{imp,\mathbf{k},\sigma}}=\frac{\pi n_{imp}|v_0|^2}{\hbar} \int\frac{d^3 k'}{(2\pi)^3}\delta(\xi_{\mathbf{k},\sigma}-\xi_{\mathbf{k}',\sigma}) \text{ .}
\label{eq:Gamma_imp}
\end{equation}

\noindent For parabolic bands, Eq.(\ref{eq:Gamma_imp}) can be calculated analytically, yielding, at the Fermi level,
\begin{equation}
\Gamma_{imp,\sigma}=n_{imp}|v_0|^2\frac{2m}{4\pi\hbar^3}k_{F,\sigma} \text{ ,}
\label{eq:Gamma_imp_parabola}
\end{equation}

\noindent where $k_{F,\sigma}$ is the Fermi-level momentum of the spin-$\sigma$ band. From Eq.(\ref{eq:Gamma_imp_parabola}), it becomes evident that the impurity scattering rate is larger for the band  that has the larger Fermi surface. In the case of hole-like bands, this means, in contrast with the magnetic scattering rate, that $\Gamma_{imp,\uparrow}<\Gamma_{imp,\downarrow}$.

\section{Supplementary Note 4. Effective magnon mass}
We mentioned before that we use a parabolic approximation for the energy dispersion of the magnons in order to calculate the magnetic scattering rate [see Eq.(\ref{eq:quadratic_bosons})]. Here, we show how we estimated the effective magnon mass $M$ for the case of EuCd$_2$As$_2$.

Expanding Eq.(\ref{eq:Omega0a}) up to second order in $\mathbf{q}\cdot\boldsymbol{\delta}$, we find
\begin{equation}
\Omega_{\mathbf{q}}^{(0)}=J_0\hbar^2S\left[(\mathbf{q}\cdot\boldsymbol{\delta}_{1})^2+(\mathbf{q}\cdot\boldsymbol{\delta}_{2})^2+(\mathbf{q}\cdot\boldsymbol{\delta}_{3})^2\right] \text{ ,}
\label{eq:Omega_expansion}
\end{equation}

\noindent where $\boldsymbol{\delta}_{i}$ ($i=1,2,3$) are the vectors connecting the first-neighbor sites of the magnetic lattice. The Eu atoms in EuCd$_2$As$_2$ form an simple hexagonal lattice, for which
\begin{align}
& \boldsymbol{\delta}_1=a\hat{x} \text{ ,}\label{eq:delta_1}\\
& \boldsymbol{\delta}_2=\frac{a}{2}\hat{x}+\frac{\sqrt{3}a}{2}\hat{y}\text{ ,}\\
& \boldsymbol{\delta}_3=c\hat{z} \label{eq:delta_3}\text{ ,}
\end{align}

\noindent with $a\approx 4.43$ \AA and $c\approx 7.32$ \AA. Substituting Eqs.(\ref{eq:delta_1})-(\ref{eq:delta_3}) into Eq.(\ref{eq:Omega_expansion}), we obtain, neglecting the cross terms
\begin{equation}
\Omega_{\mathbf{q}}^{(0)}\approx J_0\hbar^2S\left(\frac{5a^2}{4}q_x^{2}+\frac{3a^2}{4}q_y^2+c^2q_z^2\right)\text{ .}
\end{equation}

\noindent For simplicity, we further approximate he boson dispersion to be isotropic. In this case,
\begin{equation}
\Omega_{\mathbf{q}}^{(0)}\equiv J_0\hbar^2S\left\langle a^2\right\rangle =\frac{\hbar^2q^2}{2M} \text{ ,}
\end{equation}

\noindent where $\left\langle a^2 \right\rangle=(2a^2+c^2)/3\approx 3.1\times 10^{-19}$ m$^{-1}$, and therefore $M=1/(2J_0S)\left\langle a^2\right\rangle\approx 7.5\times 10^{-28}$ Kg.

\section{Supplementary Note 5. Carrier density}

 Hall conductivity data gives a carrier density of $n=2.35\times 10^{26}$ m$^{-3}$ for EuCd$_2$As$_2$, where the majority of carriers are holes. We use such carrier density to calculate the chemical potential $\tilde{\mu}=W-\mu$ self-consistently trough $N_h=N_{tot}-N_e$, where $N_{h}$ is the number of the holes in the system, which is obtained by subtracting from the total number of states that fits in a parabolic band with band-width $\Lambda$,
\begin{equation}
N_{tot}=N^{(tot)}_{\uparrow}+N^{(tot)}_{\downarrow}=\frac{V}{4\pi^2}\frac{(2m)^{3/2}}{\hbar^3}\left[\int\limits_{-\Lambda}^{\tilde{\mu}-\gamma}d\tilde{\xi}\sqrt{\tilde{\mu}-\gamma-\tilde{\xi}}+\int\limits_{-\Lambda}^{\tilde{\mu}+\gamma}d\tilde{\xi}\sqrt{\tilde{\mu}+\gamma-\tilde{\xi}}\right] \text{ ,}
\label{eq:Ntot}
\end{equation}

\noindent from the total number of electrons
\begin{equation}
N_{e}=N_{e,\uparrow}+N_{e,\downarrow}=\frac{V}{4\pi^2}\frac{(2m)^{3/2}}{\hbar^3}\left[\int\limits_{-\Lambda}^{\tilde{\mu}-\gamma}d\tilde{\xi}\sqrt{\tilde{\mu}-\gamma-\tilde{\xi}}\frac{1}{e^{\tilde{\xi}/k_B T}+1}+\int\limits_{-\Lambda}^{\tilde{\mu}+\gamma}d\tilde{\xi}\sqrt{\tilde{\mu}+\gamma-\tilde{\xi}}\frac{1}{e^{\tilde{\xi}/k_B T}+1}\right] \text{ .}
\label{eq:Ne}
\end{equation}

\noindent In Eqs.(\ref{eq:Ntot}) and (\ref{eq:Ne}), $\gamma=\gamma(T)$ defined in Eq.(\ref{eq:xi_sigma}). The resulting density of holes $n_h=N_h/V$ is independent of the band width, as long as $\Lambda\gg \tilde{\mu}\pm\gamma$.

\section{Supplementary Note 6. Complementary figure}

\begin{figure}[t!]
\centering
\includegraphics[width=0.6\textwidth]{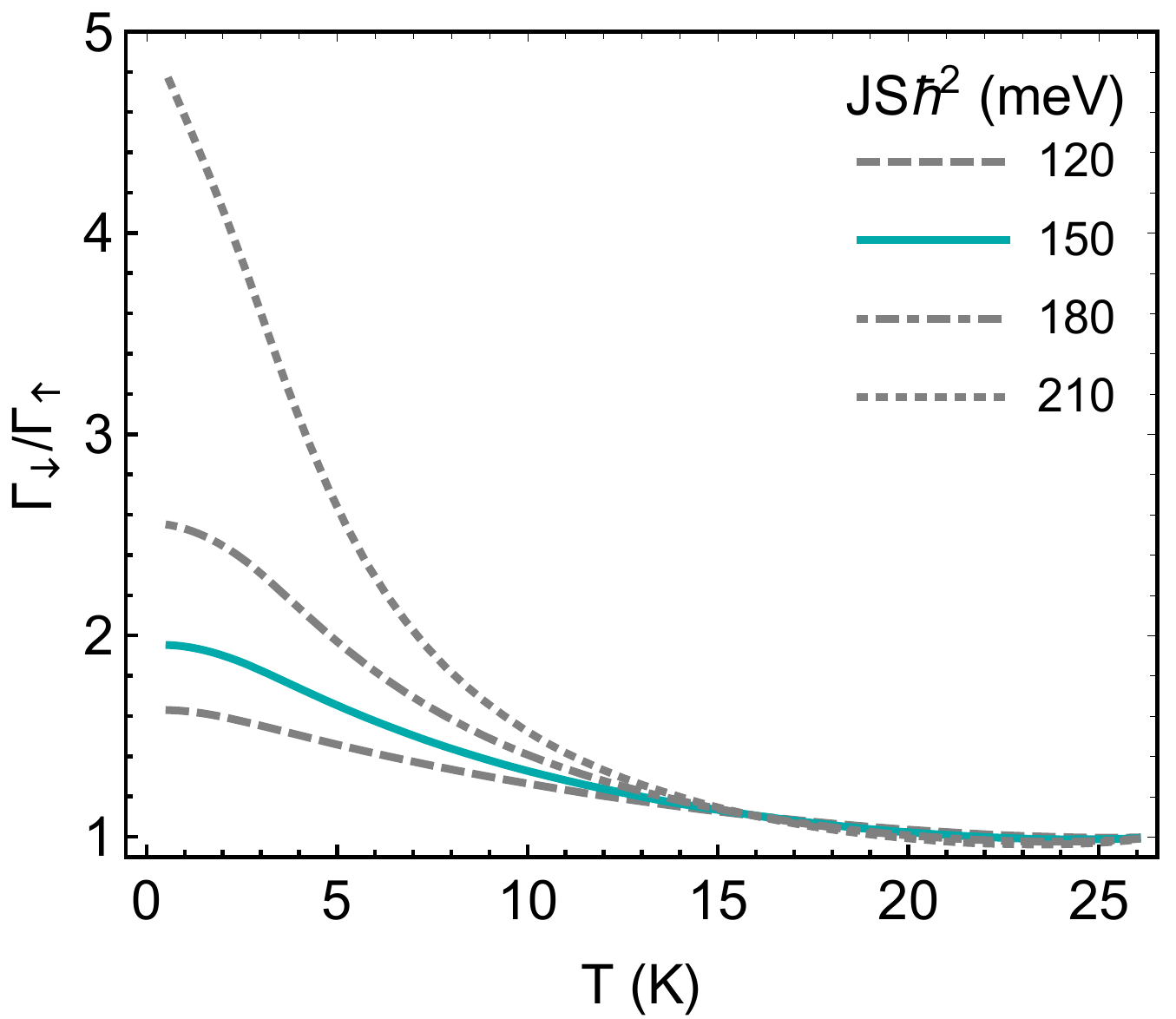}
\caption{Ratio of the total scattering rates $\Gamma_{\downarrow}/\Gamma_{\uparrow}$ as function of temperature for different values of $J$. We set $n=n_0=2.35\times 10^{26}m^{-3}$ and $J_0S\hbar^2=0.15meV$. The values of $J$ were carefully chosen so that both majority and minority bands crosses the Fermi level.}
\label{fig:Dyson_disorder}
\end{figure}

We showed that the impurity scattering dominates over magnetic scattering at low temperatures in EuCd$_2$As$_2$. As a consequence, $\Gamma_{\downarrow}>\Gamma_{\uparrow}$ (recalling that $\Gamma_{\sigma}=\Gamma_{mag,\sigma}+\Gamma_{imp,\sigma}$) for small $T$, as observed in ARPES data. In the figures shown in the main text of this manuscript, we set $JS\hbar^2=150 meV$, which we estimated using the experimentally observed splitting of minority and majority bands. However, there are other effects, such as lattice deformation, that could contribute to the band shift. This gives a natural uncertainty for the estimated value of $J$, and to complement our results, we show here the ratio $\Gamma_{\downarrow}/\Gamma_{\uparrow}$ as a function of temperature for different values of $J$. We see that the sharpening of the inner band is more pronounced for larger values of the coupling between the itinerant electron and the localized moments.

\end{document}